# Uncertainty Principle:
# Classic and Quantum Aspects

Brazovskaja N.V, Brazovsky V.Ye. (braz@agtu.secna.ru)
General Physics Dept, Altai State Technical University, Barnaul, Russia.

**Abstract**. Some aspects of application of the Uncertainty Principle in the range of interaction radiation with matter surveyed. The procedure of adjustment is proposed at calculation of values of an electromagnetic energy in a quantum theory of a field.

## 1. Introduction

The electromagnetic field has appeared by that rock, about which one a classic mechanics - in a problem on a caloradiance in due course has stumbled. The creation of lasers with a major radiant density - such field, it is doubtless, it is necessary to consider as classic - puts a number of problems, to refer which one only to classic or only to a quantum theory it appears inconvenient. In the present paper we attempt to analyze some aspects of an indeterminacy relation with reference to an impulse electromagnetic radiation and we shall make some generalizations.

## 2. Classic theory

The concept of a photon can be introduced from positions of the classic theory as physical object, particle, some impulse, emitted by atom in one act of transition between two energy levels.

1. We shall esteem a phrase "atom radiates a photon with energy $\hbar\omega$, equal difference of energies of levels of electronic transition". Let's begin that of a photon with just the same energy, i.e. $E=\hbar\omega$ can not basically exist. Let's clarify why, and that exists or can exist actually. Here the word "actually" means "within the framework of the existing physical theory".

Classic model. Let atom radiates during time $\tau$ a segment of a sinusoid:

$$A=A_0 \cos(\omega t - \mathbf{kr}). \qquad (1)$$

The frequency distribution in our case is given by expression (we use complex representation):

$$I(\omega) = \left| \int_0^\tau A_0 \exp(\omega_0 t) \exp(-i\omega t) dt \right|^2 = \frac{A_0^2}{(\omega-\omega_0)^2} \sin^2\left[\frac{(\omega-\omega_0)\tau}{2}\right], \qquad (2)$$

Indicating, that our radiation (the researched photon) contains not one frequency, and frequency spectrum with width $1/\tau$, that expresses an uncertainty relation $\Delta\omega\tau \geq 2\pi$ [1]. Let's remark at once, that the uncertainty relation is not the product of a quantum theory, and takes place always, the Fourier transform linking temporary and frequency patterns of the process is soon fair. The frequency is bound to energy by a constant factor $\hbar$, therefore our photon has not strictly particular energy, and spectrum of energies. From a quantum theory in this case there is only relation between frequency and energy.

We would seem have mated the classic and quantum approaches and in this case we work extremely with a quantum theory. For calculation of energy of the undular surveyed package it is necessary to take expression for bulk density of energy of an electromagnetic field with usage of strengths of electrical and magnetic fields and integrate on volume. Actually indicated integration can be fulfilled only for all space. The impulse timebounded - to integrate only on space disregarding boundednesses in time it is impossible. The introduction of restricted space automatically reduces in quantization

For precise value of energy our photon should exist indefinitely long or, if it is possible so to be expressed, to have infinite length. The expression can be applied for energy by the way $E=\hbar\omega$ in the event that in radiation there is only one frequency. Any radiation exists final time therefore any sub-stantial photon can have of precise value of energy, moreover, any substantial photon can have defined values of energy. It is meaningful to speak only about availability of average value of energy, characterizing a photon by allocation on energies. The given assertion in a strong sense concerns any physical object.

Extending, it is possible to state, that in a strong sense we should not characterize physical objects of energy (no less than other physical properties) as by C-number. The physical object should be characterized by distribution on energies, which one has average value and other moments. The distribution - is reflection, the characteristic of some state, in which one is the physical object, in a considered case a photon, which one we in this case consider from positions of the classic theory.

About radiation of final length usually speak as about a wave packet, i.e. substantial photon always - a wave packet, physical condition of a photon characterized by some distribution. It is possible to speak about average energy $E=\hbar\omega_0$ (first moment of distribution) and uncertainty of energy $\Delta E \sim 2\pi\hbar/\tau$ (second moment), which one uniquely are bound with appropriate by a difference of energies of atomic transition and its width.

2. About a direction of photons. Let's consider expression "the atom absorbs directional lengthwise axis Z laser photons".

That it is better to imagine an originating in this case problem, we shall imagine, that on paths of a wave directly ahead of atom we have delivered a diaphragm. Apparently already it is impossible to state, that behind a diaphragm the photons are directed lengthwise axis Z. But you see the diaphragm (theoretical) in any way does not change a direction of propagation of photons, it only "cuts" a part of a light bundle. Than the photons which have come to atom if there is a diaphragm and in its absence differ? By anything. If there is a diaphragm them it will be simple less. Therefore, the initial assertion from the given expression is not quite correct.

The essence of the set problem consists besides basically of uncertainties. Only in item 1 this relation was utillized for steams energy - time (E, t), and in the given item - for steams pulse - coordinate ($\mathbf{k}$, r).

So, formulation. In space there is a plane simple harmonic wave characterized by a wave vector $\mathbf{k}$. (Plane it can be only in case of absence of limitations in a direction, perpendicular direction of propagation.) In some point of space characterized the position vector r, poses atom, which one should absorbs a photon from a given plane wave. What pulse will receive atom, having absorbed one (only one) photon? The formulation is lawful, and is lawful from positions of the classic theory, as the Huygens-Fresnel principle concerns to the classic theory.

The indeterminacy relation in this case links variables $\mathbf{k}$ and $\mathbf{r}$. The photon should be absorbed in a point of localization of atom, i.e. should have a defined value $\mathbf{r}$ at the moment of absorption. Therefore, by virtue of a principle of uncertainty, the value of its pulse completely undeterminable. But the wave monochromatic, i.e. absolute value k = |$\mathbf{k}$| is uniquely determinated by a radiated frequency. A direction of a vector $\mathbf{k}$ in this case can be undeterminated only. The given vector has an random direction in limits of a spherical corner $2\pi$ steradian around of an axis defined by a vector $\mathbf{k}$.

### 3. Quantum theory

Let us consider in more detail following situation evocative in a cut-in: at calculation by methods of a quantum mechanics of value of different energies (difference of energies any of transition, the electrostatic energies etc.) are gained by the way E+i$\Delta$E, i.e. the calculated value of energy, as a rule, has an imaginary component [2]. But the physical quantity, according to rules of a quantum theory, should be true. What one must to do with an imaginary part? To discard or to take as the required unit of the value? Is already two ways of calculation. Where the warranty, what by exact will not appear any third way of calculation? On what basis it is necessary to select this or that operation with a calculated result?

To answer the given problem it is necessary to clarify physical sense of true and imaginary parts of the retrieved value. We already have clarified, that a particle or the interreacting particles can be in some quantum states, and these quantum states are equivalent, i.e. are selected by the way of base for decomposition, but only from a linear case. The process of interaction is basically nonlinear operation. The quantum states, in which one are interreacting particles, to us are unknown.

In the theory of radiations the imaginary part of energy contacts to $\tau^{-1}$, here $\tau$ is a lifetime of an exited state, width (uncertainty) of an energy level is entered. And as required value of energy the real part of energy is selected. So, with physical sense of an imaginary part all is clear, and the second problem remains. Whether the choice of a real part is enough justified as required value of energy?

The assertion about a reality of a gauged physical quantity in a nomenclature of quantum states, apparently, should mean following. At measurement of some physical quantity we obtain a quantum state, in which one the datum has a defined value (and all remaining values have this or that degree of uncertainty). Then logically reasonable there will be a following operation. Let as a result of calculations we have received some value of a physical quantity B ($x_1$, $x_2$, …), which one represented in complex sort. It can mean alone - we have received a physical quantity In in that quantum state (in that representation), which one is necessary for us, as we can not know this state a priori.

**The procedure**. As the uncertainty of the quantum value for us is bound to an imaginary part of expression defining a datum, shall select uncertainties of arguments $x_1$, $x_2$, … so that value In became real. Differently, the procedure is encompass byed following overdetermination of arguments:

$$x_k \to x_k + i\zeta_k \tag{3}$$

And guard rope of values $\zeta_k$ so that

$$\mathrm{Im}B(\zeta_k)=0. \tag{4}$$

For a case of one variable the equation (4) uniquely determinates the value $\zeta$, substituting which one in expression of the required value In, we discover demanded true value last. The given procedure for the first time utilised in paper [3] for calculation of energy by induced radiation of an intermolecular interaction.

The circumscribed procedure is convenient for using, if the dependence of the required value on some parameters is known. Let's esteem a case, when the dependence of some value on parameters is unknown. For example, energy of some transition in atom presets by the way $E+i\Delta E$. In this case we shall speak about obtaining an estimation of a desired value of energy.

Let's write a combination of energy with padding to it by a variable - in due course by the way:

$$(E+i\Delta E)t \tag{5}$$

Let's execute overdetermination of time, i.e. we shall enter uncertainty of time t:

$$t \to t+i\tau \tag{6}$$

The expression (5) becomes:

$$(E+i\Delta E)(t+i\tau) = Et - \Delta E\tau + i(E\tau + \Delta Et) \tag{7}$$

Let's make the equation (4):

$$(E\tau + \Delta Et)=0 \tag{8}$$

We obtain:

$$\tau = \frac{\Delta E t}{E} \tag{9}$$

Substituting in (7), we discover:

$$\left(E - \frac{(\Delta E)^2}{E}\right)t \tag{10}$$

Thus, the required value of energy is

$$E - \frac{(\Delta E)^2}{E}. \tag{11}$$

As it is uneasy to see, the obtained value of energy (in representation with particular energy) differs from a true part of expression (5). It is possible to consider, that we have received more close to true value of energy. (As analogy it is desirable to recollect difference between own and resonant frequencies in theory of damped oscillations).

## 4. Outputs

1. The uncertainty relation of the Heisenberg in a traditional statement actually is the product not quantum, but classic wave theory. In the classic theory from link of frequency (energy) and temporary patterns by means of Fourier transform follows, that for the complete characteristic of the physical object it is necessary to use not numbers, and distributions. At a jumboizing of particles, i.e. at transition to macroscopic objects the appropriate distributions tend to δ-functions. For optical impulses the transition to a macroscopic limit is impossible. The optical impulse should be characterized by a combination of distributions in temporary and frequency areas.

2. The uncertainty of some parameter in a quantum theory should be connected to an imaginary part of expression, circumscribing the given parameter. And distributions of the given parameter, as it takes place in the classic theory, we can not know. It is necessary to suppose, that the given distributions in the obtained quantum state of a particle does not exist. The particle generally should be characterized by a quantum state characterized set of parameters and their uncertainties, i.e. set of complex values of parameters, from which one only one can be real. For transition to other quantum state it is necessary to change imaginary components of "waste" parameters. If there is interaction the quantum states are not equivalent.